\documentclass[sigconf]{acmart}

\acmConference[SC '25]{The International Conference for High Performance Computing, Networking, Storage and Analysis}{November 16--21, 2025}{St Louis, MO, USA}
\acmBooktitle{The International Conference for High Performance Computing, Networking, Storage and Analysis (SC '25), November 16--21, 2025, St Louis, MO, USA}

\usepackage{graphicx}
\usepackage{subcaption}
\usepackage{cleveref}
\usepackage{makecell}
\usepackage{amsmath}
\usepackage{multirow}
\usepackage{xcolor}
\usepackage{fancyvrb}
\usepackage{tikz}
\usepackage{cprotect}
\usepackage{caption}

\DeclareMathOperator*{\argmax}{argmax}

\DefineVerbatimEnvironment{code}{Verbatim}{fontsize=\small}

\definecolor{lightblue}{rgb}{0.6509803921568628,0.807843137254902,0.8901960784313725}
\definecolor{darkblue}{rgb}{0.12156862745098039,0.47058823529411764,0.7058823529411765}
\definecolor{lightgreen}{rgb}{0.6980392156862745,0.8745098039215686,0.5411764705882353}
\definecolor{darkgreen}{rgb}{0.2,0.6274509803921569,0.17254901960784313}
\definecolor{lightred}{rgb}{0.984313725490196,0.6039215686274509,0.6}
\definecolor{darkred}{rgb}{0.8901960784313725,0.10196078431372549,0.10980392156862745}
\definecolor{lightorange}{rgb}{0.9921568627450981,0.7490196078431373,0.43529411764705883}
\definecolor{darkorange}{rgb}{1.0,0.4980392156862745,0.0}
\definecolor{lightviolet}{rgb}{0.792156862745098,0.6980392156862745,0.8392156862745098}
\definecolor{darkviolet}{rgb}{0.41568627450980394,0.23921568627450981,0.6039215686274509}
\definecolor{yellow}{rgb}{1.0,1.0,0.6}
\definecolor{brown}{rgb}{0.6941176470588235,0.34901960784313724,0.1568627450980392}
\begin{document}

\title{PerfDojo: Automated ML Library Generation for Heterogeneous Architectures}

\author{Andrei Ivanov}
\affiliation{%
  \institution{ETH Zurich}
  \city{Zurich}
  \country{Switzerland}
}
\email{andrei.ivanov@inf.ethz.ch}

\author{Siyuan Shen}
\affiliation{%
  \institution{ETH Zurich}
  \city{Zurich}
  \country{Switzerland}
}

\author{Gioele Gottardo}
\affiliation{%
  \institution{ETH Zurich}
  \city{Zurich}
  \country{Switzerland}
}

\author{Marcin Chrapek}
\affiliation{%
  \institution{ETH Zurich}
  \city{Zurich}
  \country{Switzerland}
}

\author{Afif Boudaoud}
\affiliation{%
  \institution{ETH Zurich}
  \city{Zurich}
  \country{Switzerland}
}

\author{Timo Schneider}
\affiliation{%
  \institution{ETH Zurich}
  \city{Zurich}
  \country{Switzerland}
}

\author{Luca Benini}
\affiliation{%
  \institution{ETH Zurich}
  \city{Zurich}
  \country{Switzerland}
}

\author{Torsten Hoefler}
\affiliation{%
  \institution{ETH Zurich}
  \city{Zurich}
  \country{Switzerland}
}

\renewcommand{\shortauthors}{Ivanov et al.}

\begin{abstract}
  The increasing complexity of machine learning models and the proliferation of diverse hardware architectures (CPUs, GPUs, accelerators) make achieving optimal performance a significant challenge. Heterogeneity in instruction sets, specialized kernel requirements for different data types and model features (e.g., sparsity, quantization), and architecture-specific optimizations complicate performance tuning. Manual optimization is resource-intensive, while existing automatic approaches often rely on complex hardware-specific heuristics and uninterpretable intermediate representations, hindering performance portability. We introduce PerfLLM, a novel automatic optimization methodology leveraging Large Language Models (LLMs) and Reinforcement Learning (RL). Central to this is PerfDojo, an environment framing optimization as an RL game using a human-readable, mathematically-inspired code representation that guarantees semantic validity through transformations. This allows effective optimization without prior hardware knowledge, facilitating both human analysis and RL agent training. We demonstrate PerfLLM's ability to achieve significant performance gains across diverse CPU (x86, Arm, RISC-V) and GPU architectures.
\end{abstract}


\begin{teaserfigure}
    \centering
    \begin{subfigure}{0.5\linewidth}
        \centering
        \includegraphics[width=1\linewidth]{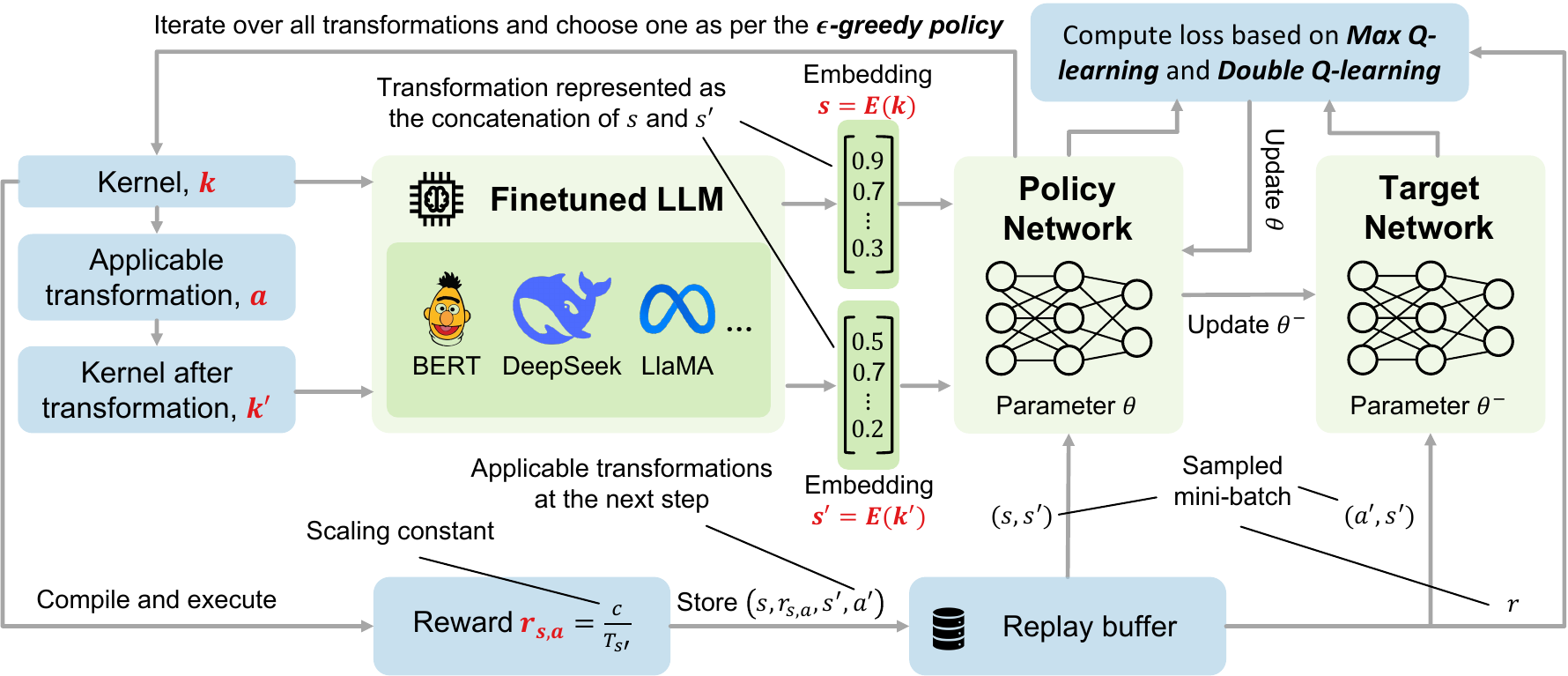}
        \caption{}
        \label{fig:perfllm-overview}
    \end{subfigure}
    \begin{subfigure}{0.46\linewidth}
    \centering
    \includegraphics[width=1\linewidth]{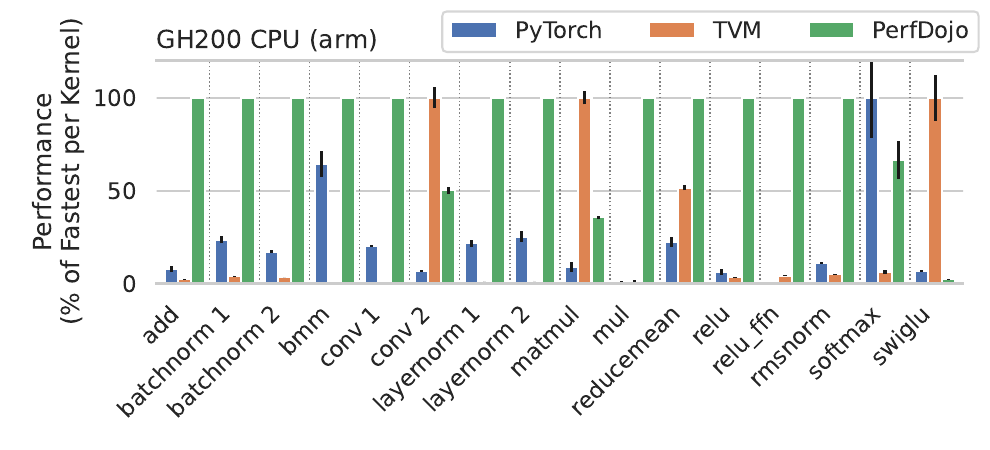}
    \caption{}
    \label{fig:speedup-gh200-cpu}
    \end{subfigure}
    \caption{(a) Overview of the PerfLLM training pipeline. (b) On GH200, the geometric mean speedup of PerfDojo is 6.65$\times$ relative to PyTorch~\cite{paszke2019pytorch} and 13.65$\times$ relative to TVM~\cite{chen2018tvm}. The figure displays multiple variants for some kernels, reflecting the consideration of different input tensor shapes (\Cref{tab:kernels}).}
    \label{fig:poster}
\end{teaserfigure}



\maketitle

\section{Introduction}

\begin{figure*}
  \includegraphics[width=\textwidth]{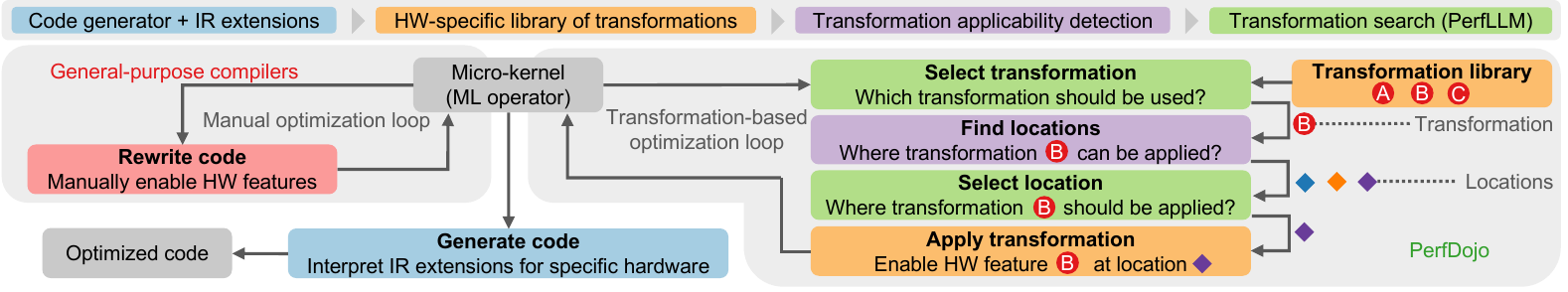}
  \caption{Manual vs. PerfDojo's transformation-centric optimization workflows.}
  \label{fig:overview}
\end{figure*}

The growing size~\cite{villalobos2022machine} of machine learning models is increasing computational demands and driving the development of a wide range of model architectures, high-performance accelerators (NVIDIA A100~\cite{a100_nvidia}, Google TPU v4~\cite{tpu_google}, Xilinx Versal FPGAs~\cite{xilinx_versal}, Snitch~\cite{zaruba2020snitch}), and processor architectures (ARM~\cite{wang2023armv8}, RISC-V~\cite{riscv_kalapothas2023survey}). Fully utilizing the performance of these systems is becoming increasingly difficult.

Achieving optimal performance utilization is complex due to the heterogeneity of accelerator architectures and their instruction sets. Each arithmetic primitive within machine learning necessitates specialized kernels for efficient computation. These kernels must be tailored to specific data types, contingent on the underlying model. Furthermore, models employing quantization or sparsity require implementation adjustments to accommodate the algorithmic subtleties of these features~\cite{frantar2025marlin}. In certain scenarios, such as those leveraging flash attention, the selection of specific models and accelerators may enable operator fusion, necessitating bespoke kernel implementations~\cite{shah2024flashattention}.

Despite variations in implementation, common optimization strategies, such as tiling, vectorization, padding, and the utilization of specialized instructions, remain prevalent. However, these optimizations necessitate architecture-specific parameterization, accounting for memory and cache sizes, available specialized instructions, and other hardware attributes. Moreover, the code syntax, exemplified by CUDA~\cite{luebke2008cuda}, is vendor-dependent and dictated by available compilation stacks.

Optimizing for all these architectures is challenging. To address them efficiently, we propose a novel automatic optimization approach relying on LLMs. Central to enabling our method is PerfDojo, an environment that frames code optimization as an RL~\cite{sutton2018reinforcement} game. At its core lies a flexible, human-readable representation of programs and their transformations, closely resembling mathematical formulas accompanied by performance annotations that clarify their mapping onto hardware features. Inspired by recent LLM architectures~\cite{guo2025deepseek,besta2024graph}, PerfDojo is designed with modularity and interpretability at its heart. This human-oriented design not only facilitates manual optimization but also enables RL agents to explore and apply code transformations more effectively.

Manual optimization is expensive, demanding significant man-hours to acquire the necessary architectural expertise. Even with tools assisting in the optimization process (e.g. DaCe~\cite{ben2019stateful}), achieving effective results remains challenging. These tools often require a deep understanding of hardware constraints to select appropriate optimizations. For instance, choosing the correct loop unrolling factors or memory access patterns still necessitates a thorough grasp of the target architecture, despite tool assistance.

Existing automatic optimization frameworks often fail to match the performance of hand-tuned libraries due to several key limitations. First, they rely on hardware-aware heuristics to guide optimization searches. This requires vendors to implement complex heuristics that represent specific hardware knowledge. Second, these frameworks impose a significant burden on vendors of new hardware architectures and framework developers, who must maintain specialized code generator backends. Finally, these frameworks lack an API for automating the fine-grained code transformation workflow that human engineers use during manual code optimization.

In this work, we propose PerfDojo (\Cref{fig:overview}), a code representation and environment for code transformation prioritizing the validity of code transformations. This approach ensures that the automatic exploration of the program optimization space is restricted to programs with unaltered semantics.
\begin{itemize}
    \item Specifically, PerfDojo allows fully automated code optimization without explicit specification of hardware models or heuristics, and it provides guarantees on program semantic preservation necessary for the full automation of the code optimization process.
    \item Moreover, the representation is engineered for human readability and interpretability. This is not merely a convenience, but a strategic necessity. Human understanding facilitates effective debugging, iterative refinement, and the development of intuitive heuristics, all of which are essential for constructing a robust learning environment. The ability of humans to effectively interact with and understand the representation directly translates to the potential for RL agents to learn and optimize within it.
    \item Finally, PerfDojo shifts the paradigm from providing hard\-ware-aware libraries to providing hardware-aware transformations, enabling vendors to extend the automatic library generation pipeline.
\end{itemize}

Our key contributions involve:
\begin{itemize}
    \item Formulating PerfDojo: A novel representation guaranteeing semantic validity preservation throughout automatic code transformations (\Cref{sec:perfdojo}).
    \item Formulating PerfLLM: A novel automatic program optimization methodology leveraging LLMs and RL, optimizing code without prior architecture knowledge (\Cref{sec:perfllm}).
    \item Evaluating achievable performance gains with our representation and optimization methodology on a set of ML kernels and hardware architectures (\Cref{sec:evaluation}).
\end{itemize}

\section{PerfDojo: A Game for Finding High-Performance Kernels}
\label{sec:perfdojo}

Creating a Dojo for RL agents is not trivial. Firstly, it needs to support implementations that could be handcrafted by human developers. Secondly, there should be streamlined progression toward these implementations.

One feature of the representation we require is support for manual optimization processes that can be performed by a human engineer. General-purpose compilers, such as Clang and GCC, lack fine-grained control over the application of transformations to user-specified sections of the intermediate representation. This limitation restricts the search space to cases predefined by heuristics built into the compilers, and it complicates the exposure of transformation-centric optimization pipelines to search methods that are not integrated into the compilers.

Some frameworks may allow for detailed specification of transformation schedules without checking correctness of these, making it the responsibility of a user to make sure that requested transformations are not breaking code semantics. While this may be acceptable for manual optimization, the lack of thorough applicability checks accompanied with transformations make search space of program optimizations filled with many broken implementations. Although this can be remedied with numerical verification, it will pollute the search space requiring longer exploration time by automatic optimization methods. We expect that enforcing semantic preservation in the IR manually will help RL agents to focus only on the program optimization task, instead of trying additionally to learn how to avoid semantically incorrect transformations.

A feature helpful for maintaining interpretability of IR is keeping each transformation simple, that is, atomically doing only one specific change at once. This can ensure that programmers can verify the result of this change manually, making the transformation semantics also understandable for RL agents.

Simple transformations can help formulate straightforward applicability constraints for the transformation under which semantic preservation is guaranteed, which is very challenging for non-atomic transformations, as demonstrated in FuzzyFlow~\cite{schaad2023fuzzyflow}. In some frameworks, loop vectorization can be an example of a complex transformation, when it handleds loop tiling to the vector size, followed by unrolling of the innermost loop and attempting to replace each vectorizable operand and instruction in its unrolled body with appropriate vector counterpart. In contrast, PerfDojo requires tiling transformation to be applied explicitly, only after which vectorization can be applied under the constraint that the number of iterations in the affected loop are equal to the vector size and this loop wraps just a single instruction with vectorizable arguments. 

When the program optimization search space is getting too large, existing frameworks usually resort to incorporating heuristics that guide transformations according to expert knowledge. Collecting such knowledge and describing it in a heuristic may be a very cumbersome process, limiting the speed with which these frameworks are adjusted to the new hardware features. 

Sometimes, for the sake of simplicity, search space may be explicitly constrained beyond constraints imposed by representation (e.g. to search only over tile sizes) either because it is simpler to ensure semantic preservation for a limited subset of programs or to constrain search space with heuristics to avoid long transformation search times. Such constraints may disable finding of the most efficient implementations.  

We also ask for transformations to be non-destructive, meaning they do not lose any details of the initial computation specification. For example, if loop unrolling decomposes each loop iteration to individual instruction, it may be impossible to undo it. Both human engineers and RL agents may become aware of incorrectly applied unrolling in the earlier transformation step and may want to undo it, maintaining all other transformations applied since then in place.

\begin{table}
  \caption{Features Available in Representations of Existing Frameworks}
  \label{tab:representation-features}
  \small
  \begin{tabular}{@{}l@{ }r@{ }r@{ }r@{ }r@{ }r@{ }r@{}}
    \toprule
    & GCC & Polly & Halide & Dace & TVM & PerfDojo \\
    \midrule
    \makecell[l]{Manual transformations} & $\times$ & $\times$ & $\checkmark$ & $\checkmark$ & $\checkmark$ & $\checkmark$ \\
    \makecell[l]{Semantic preservation} & $\checkmark$ & $\checkmark$ & $\times$ & $\times$ & $\checkmark$ & $\checkmark$ \\
    \makecell[l]{Atomic transformations} & $\times$ & $\times$ & $\times$ & $\times$ & $\checkmark$ & $\checkmark$ \\
    \makecell[l]{Heuristics not required} & $\times$ & $\times$ & $\checkmark$ & $\checkmark$ & $\times$ & $\checkmark$ \\
    \makecell[l]{Unconstrained search space} & $\times$ & $\checkmark$ & $\times$ & $\checkmark$ & $\times$ & $\checkmark$ \\
    \makecell[l]{Non-destructive transformations} & $\times$ & $\checkmark$ & $\times$ & $\times$ & $\times$ & $\checkmark$ \\
  \bottomrule
\end{tabular}
\end{table}

In \Cref{tab:representation-features}, we list frameworks along with the requirements their representations satisfy. In the following sections, we describe a representation and transformation space designed to satisfy these requirements.

\subsection{Representation}
\label{sec:representation}

\begin{figure}
    \centering
    \begin{subfigure}{1\linewidth}
        \centering
        \begin{subfigure}{0.42\linewidth}
            \includegraphics[trim={0 5 10 7},clip,width=\linewidth]{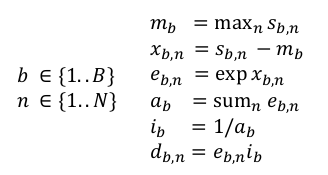}
            \caption{Math notation}
        \end{subfigure}
        \begin{subfigure}{0.55\linewidth}
            \includegraphics[trim={7 7 10 6},clip,width=\linewidth]{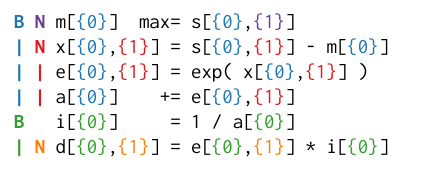}
            \caption{Textual description}
            \label{fig:textual-description}
        \end{subfigure}
        \begin{subfigure}{0.8\linewidth}
            \includegraphics[trim={0 0 4 4},clip,width=\linewidth]{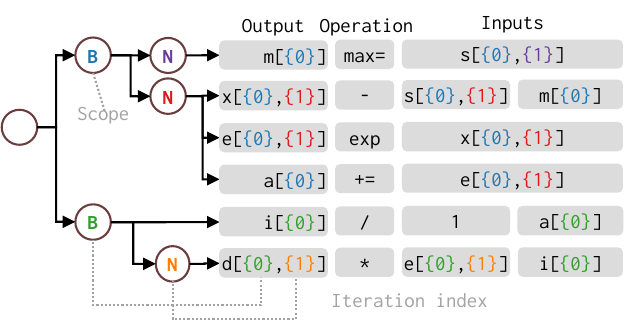}
            \caption{Graph representation}
            \label{fig:graph-ir}
        \end{subfigure}
    \end{subfigure}
    \begin{subfigure}{1\linewidth}
        \includegraphics[trim={4 4 4 4},clip,width=\linewidth]{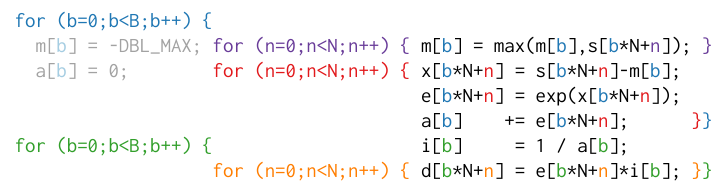}
        \caption{Generated code}
    \end{subfigure}
    \caption{Softmax kernel representations.}
    \label{fig:ir-example}
\end{figure}


\begin{figure*}
  \includegraphics[width=\textwidth]{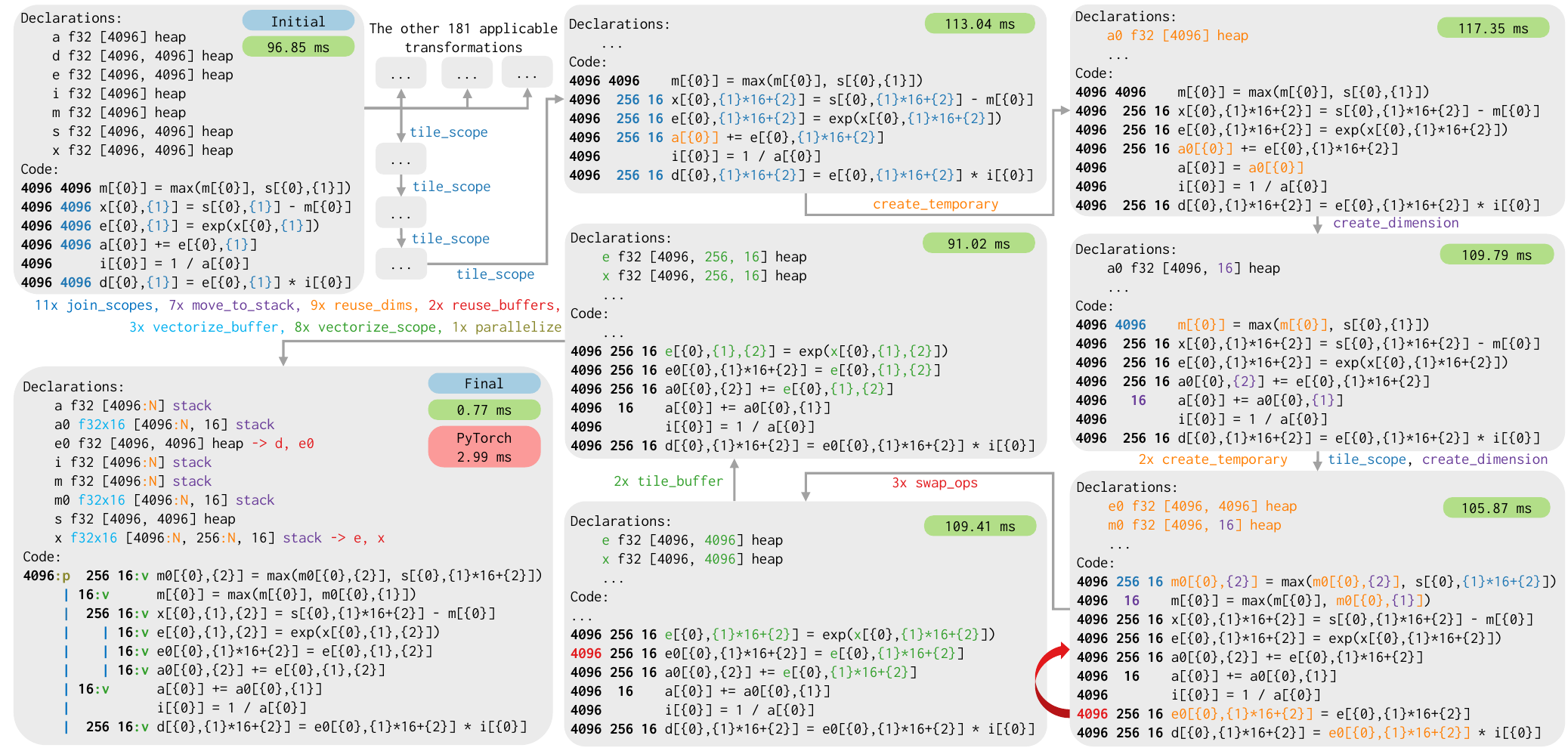}
  \caption{Optimization of a softmax kernel through a sequence of transformations (moves) on a CPU with AVX-512 extensions. Each move in the PerfDojo game maintains the initial program semantics.}
  \label{fig:softmax-transformations}
\end{figure*}

\begin{figure}
    \centering
    \includegraphics[width=1\linewidth]{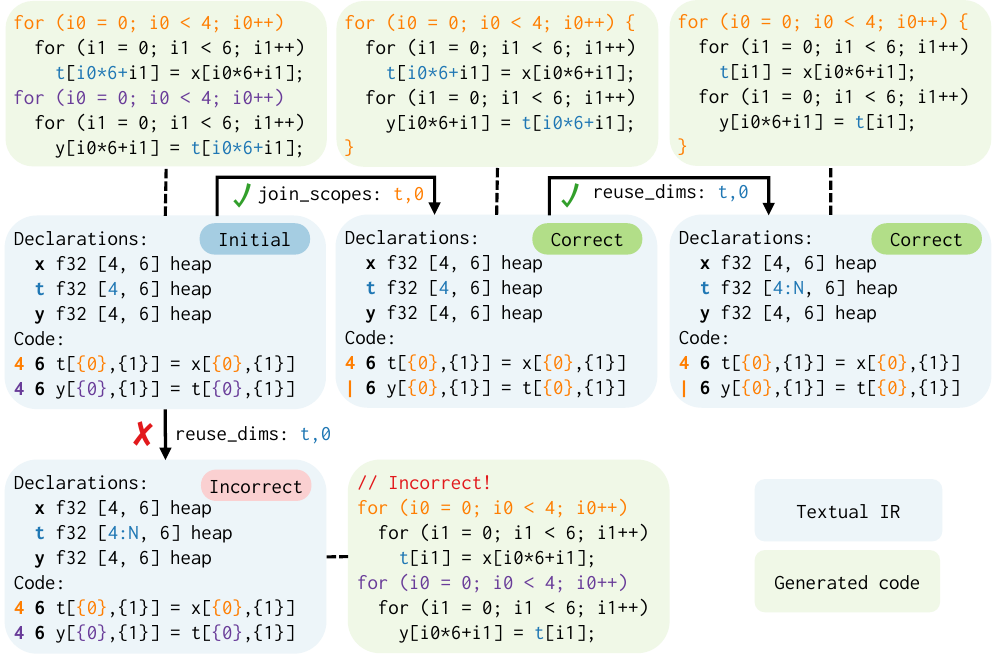}
    \caption{Program transformation example: Buffer dimension reuse (\texttt{reuse\_dims}) is correctly applied with prior loop fusion (\texttt{join\_scopes}), as shown in the top, but yields incorrect computation without it, as shown in the bottom.}
    \label{fig:detailed-transformations}
\end{figure}

The main data structure of our IR (\Cref{fig:ir-example}) is an ordered tree. Internal vertices, except the root node, represent single-dimensional iteration \emph{scopes}. Leaves represent \emph{operations}, together with associated output and inputs. The order of child vertices defines the order of their execution within the parent. Each input and output within the leaf represents a scalar value, encoded with the name of the multidimensional array and the multidimensional index within such array. Such an index can refer to the iteration over a particular ancestor scope by specifying an integer in curly braces that indicates the depth of the scope.

To facilitate operations on the graph intermediate representation (\Cref{fig:graph-ir}), we map it to a human-readable textual format (\Cref{fig:textual-description}). In this format, vertical bars \Verb@|@ denote a child node relationship with the nearest preceding line that does not contain a bar in the same position, indicating the parent node.

To specify the memory mapping of multidimensional arrays used in the textual description of the kernel, we include a list of buffer declarations. Each declaration specifies the buffer's name, the data type it holds, its shape as a list of dimensions (\Verb|[dim1, dim2]|), and its memory location (e.g., heap or stack), followed by an optional list of array names when multiple arrays reside in the same memory buffer. A single buffer declaration adheres to the following format:

\begin{code}
buffer_name data_type shape location -> list_of_array_names
\end{code}

Each dimension can have an optional \Verb|:N| suffix, which prevents its materialization and reduces the buffer's memory footprint when the iteration order allows for memory reuse. When the list of array names is omitted, it is assumed that the buffer holds a single array of the same name.

The scope sizes in the textual description support a range of suffixes that specify the way of instantiating the iteration range: \Verb|:u| to unroll the scope, \Verb|:p| to parallelize it, and \Verb|:v| to vectorize. For kernels optimized for execution on GPUs, the suffixes \Verb|:g|, \Verb|:b|, and \Verb|:w| represent scope mapping to GPU grid, block, and warp, respectively.

\begin{table}
\caption{Supported representation features in PerfDojo.}
\small
\begin{center}
\begin{tabular}{@{}l@{\hspace{3pt}}l@{\hspace{3pt}}l@{}} 
\toprule
& Operation type & Example in textual form \\
\midrule 
\multirow{6}{*}{\rotatebox{90}{Supported}} & Element-wise & \texttt{\textbf{N M} z[\{0\},\{1\}]=x[\{0\},\{1\}]*y[\{0\},\{1\}]}
\\
& Broadcast & \texttt{\textbf{N M} z[\{0\},\textcolor{darkgreen}{\{1\}}]=x[\{0\}]} \\
& Constant as value & \texttt{\textbf{N M} z[\{0\},\{1\}]=x[\{0\},\{1\}]*\textcolor{darkgreen}{C}} \\
& Index as value & \texttt{\textbf{N M} z[\{0\},\{1\}]=x[\{0\},\{1\}]*\textcolor{darkgreen}{\{0\}}} \\
& Reduction & \texttt{\textbf{N M} z[\{0\}]+=x[\{0\},\textcolor{darkgreen}{\{1\}}]} \\
& Expression as location & \texttt{\textbf{N M} z[\{0\},\{1\}]=x[\textcolor{darkgreen}{\{0\}*N+\{1\}}]} \\
\midrule
\multirow{4}{*}{\rotatebox{90}{Excluded}} & Indirection & \texttt{\textbf{N} z[\{0\}]=x[\textcolor{darkred}{y[\{0\}]}]} \\
& Data-dependent range & \texttt{\textbf{N M[\textcolor{darkred}{\{0\}}]} z[\{0\},\{1\}]=x[\{0\},\{1\}]} \\
& Dependent iteration & \texttt{\textbf{N} z[\{0\}]=z[\{0\}\textcolor{darkred}{-1}]*y[\{0\}]} \\
& General control flow &
\texttt{\textbf{\textcolor{darkred}{while z[\{0\}++]}} z[\{0\}]=x[\{0\}]*y[\{0\}]} \\
\bottomrule
\end{tabular}
\end{center}
\label{tab:ir-features}
\end{table}

In \Cref{tab:ir-features}, we demonstrate various operation types, or features, that add expressiveness to the IR. Some of these features, such as \emph{broadcast}, \emph{constant as value}, \emph{index as value}, and \emph{reduction}, are mutually orthogonal. Importantly, certain features can be expressed by combining others; for example, \emph{indirection} along with \emph{index as value} can represent \emph{expression as location} by computing an expression, storing it in a temporary location, and then retrieving the computed value via indexing. The supported features facilitate the implementation of 83\% of the kernels defined in the ONNX specification. However, features like \emph{indirection}, \emph{data-dependent range}, \emph{dependent iteration}, and \emph{general control flow}, while offering further enhancements to IR expressiveness, were deliberately excluded due to the significant challenges they pose in maintaining semantic preservation.

\subsection{Transformations}

In \Cref{fig:detailed-transformations}, we demonstrate the effect of transformations on the generated code. Unlike traditional compiler passes, transformations in PerfDojo operate on individual locations within the code. To resolve ambiguity when multiple locations are available (e.g., there can be multiple unrelated scopes that can be fused with \Verb|join_scopes|), transformations must be supplied with a unique reference to the specific code location where they should be applied. In the provided example, \Verb|join_scopes| identifies the target scope by its depth \textcolor{darkorange}{\texttt{0}} within the scope nest that encloses the operation with the output array name \textcolor{darkorange}{\texttt{t}}. Then, it fuses this target scope with the scope that immediately follows it. \Verb|reuse_dims| identifies the affected code location based on the buffer name \textcolor{darkblue}{\texttt{t}} and the dimension at index \textcolor{darkblue}{\texttt{0}}.

We make the effect of each transformation simple enough that human engineers can verify the correctness of transformations and algorithmically describe the verification process. This enables a new program optimization paradigm that eliminates the need for expert knowledge about architecture to ensure program validity. Consequently, it allows RL to learn about architecture by experimenting with transformations without requiring any prior knowledge.

To obtain the list of program locations where transformations can be applied, we equip each transformation with corresponding applicability detection functionality. In addition to finding appropriate locations for a transformation, this functionality also filters out cases that would cause semantic violations. For instance, the semantic preservation failure demonstrated in \Cref{fig:detailed-transformations} is automatically avoided by checking if the affected buffer dimension is used in more than one scope.

Our framework ensures semantic preservation by embedding correctness analyses directly within the logic for identifying transformation locations. This integrated design prevents the application of semantically invalid transformations, removing the burden of correctness verification from the user.

This guarantee is founded upon established compiler principles. The applicability of each transformation is determined by prerequisite analyses, including traditional data dependency analysis, which are encoded into the logic for identifying applicable transformations. We empirically validate the implementation of these applicability rules by numerically comparing the output of each transformed program against its original version.

In \Cref{fig:softmax-transformations}, we show the path in the program transformation graph that leads to an efficient softmax implementation. Finding such a path is, in essence, the goal of PerfDojo's game. While multiple efficient implementations may exist, reaching them is not trivial. At each implementation variant (node of the graph), there is a different set of applicable transformations, numbering in the hundreds, and only one of them should be selected. As demonstrated in the example, a total of 56 transformations were required to reach this particular implementation.

One of the key features requested for the PerfDojo design is the presence of efficient implementations that are reachable by humans through a manual code optimization process. To verify that our transformation space design includes these implementations, we evaluate the performance of kernel implementations obtained through a manual code optimization process and compare them against state-of-the-art libraries in \Cref{fig:figure_perf_libraries}.

Overall, PerfDojo implements a comprehensive suite of common optimizations, including loop and instruction reordering, loop fusion, and vectorization. It also supports loop parallelization for CPU and GPU architectures and enables memory layout transformations such as dimension reordering, padding, and storage type selection. Additionally, it supports experimental hardware features of the Snitch architecture (\Cref{sec:snitch-performance}), such as Stream Semantic Registers (SSR) and Floating-Point Repetition (FREP).


\section{PerfLLM: Learning the Performance Game}
\label{sec:perfllm}

Motivated by the design and inherent capabilities of our intermediate representation (IR), and inspired by recent works that successfully apply reinforcement learning (RL) to code optimization~\cite{zheng2020felxtensor, cummins2021compilergym, wang2022automating}, we recognized that RL offers an effective approach to navigate the vast combinatorial search spaces generated by the number of available transformations. The sheer number of possible transformation sequences at each step presents a significant challenge, one that RL is particularly well-suited to tackle. In this section, we outline the core principles of RL and detail how we formulate the performance game as an RL problem to efficiently discover high-performance optimizations and transformations.

\paragraph{Markov Decision Process}

Most reinforcement learning (RL) problems can be modeled as Markov Decision Processes (MDPs), which capture environments where outcomes are determined by both stochastic dynamics and an agent's decisions. An MDP is defined by a \emph{state space} ($\mathcal{S}$), an \emph{action space} ($\mathcal{A}$), a state transition function ($p$), a \emph{reward function} ($r$), a discount factor ($\gamma$), and a policy ($\pi$)~\cite{woergoetter2008rl, sutton2018reinforcement}. When an agent interacts with an MDP, a sequence of trajectory can be obtained \cite{sutton2018reinforcement}: $s_0, a_0, r_1, s_1, a_1, r_2, \ldots$, where the subscript $t$ denotes the current time step, $s_t \in \mathcal{S}$, $s_t \in \mathcal{A}$. The \emph{return} after time-step $t$ can be calculated as the sum of cumulative discounted rewards that are received till the termination time $T$:
$R_t = \sum_{k=0}^{T} \gamma^k r_{t + k + 1}$.
Under the Markov property, the agent's goal becomes maximizing $R_t$. This objective is formalized through the \emph{Bellman optimality equations}, which recursively relate the optimal \emph{state-value function}, 
\begin{equation}
V^*(s) = \max_{a \in \mathcal{A}} \sum_{s' \in \mathcal{S},r} p(s',r|s,a) \left[ r + \gamma V^*(s') \right],
\end{equation}
and the optimal \emph{state-action value function},
\begin{equation}
Q^*(s,a) = \sum_{s' \in \mathcal{S},r} p(s',r|s,a) \left[ r + \gamma \max_{a' \in \mathcal{A}} Q^*(s', a') \right].
\end{equation}
These equations form the basis for iterative methods like value iteration and policy iteration, enabling the derivation of the optimal policy $\pi^*$ via the equation
\begin{equation}
\pi^*(s) = \argmax_{a \in \mathcal{A}} Q^*(s,a).
\end{equation}

The \emph{$\epsilon$-greedy policy} is a simple yet effective method for balancing exploration and exploitation in reinforcement learning. Under this approach, the agent selects the best-known action (the "greedy" action) with a high probability—typically \(1 - \epsilon\) plus a small share divided among all actions—and chooses a random action with a probability of \(\epsilon\). This strategy encourages the agent to explore new actions, especially during early training when \(\epsilon\) is set high, and gradually shifts towards exploiting known good actions as \(\epsilon\) decays, thereby facilitating faster convergence to an optimal policy.

\subsection{RL Formulation}

When formulating real-world problems as reinforcement learning tasks, the design of the state space, action space, and reward function is essential.

In PerfLLM (\Cref{fig:perfllm-overview}), the primary role of LLM is to encode the PerfDojo program representation into a numerical embedding vector. This embedding captures the program's state to provide a compact, high-level representation of the corresponding kernel configuration, eliminating the need for manual feature engineering. Mathematically, if the kernel at time step \(t\) is denoted by \(k_t\), the corresponding state is given by \(s_t=E(k_t)\), where \(E(\cdot)\) represents the embedding function.

Designing the action space presents a unique challenge due to the large, and dynamically changing, set of available transformations. To address this, we represent each action as the concatenation of the embedding of a kernel before a transformation and the embedding after the transformation is applied. In this framework, the \emph{stop} action is naturally defined by concatenating two identical embeddings, indicating no change in state.

Constructing a suitable reward function also proved complex. Our initial approach, which was based on speedup relative to the kernel's prior state, led to undesirable behaviors, such as overly conservative transformation choices or exploitation of the reward mechanism through cyclic performance degradation and recovery. To address these issues, we defined the reward as $r = \frac{c}{T_{s_i}}$, where $c$ is a scaling constant and $T_{s_i}$ is the runtime of the kernel after a transformation. This formulation incentivizes actions that improve the kernel's performance without relying on some absolute runtime as a reference. Furthermore, by providing rewards after each transformation rather than only at the end of an entire optimization episode, we mitigate problems caused by sparse rewards.

\subsection{Deep Q-Learning}

Q-learning is a foundational, model-free reinforcement learning algorithm that iteratively updates an action-value function $Q(s,a)$ to learn an optimal policy~\cite{watkins1992qlearning}. It does so by refining its estimates of expected returns based on observed rewards and the maximum future Q-value. In tabular settings, each state-action pair is maintained explicitly, but this becomes infeasible for large or continuous state spaces. Deep Q-learning (DQN) addresses this limitation by using a neural network to approximate the Q-function, enabling an agent to learn effectively even when the state space is too large to enumerate~\cite{mnih2013dqn}.

An alternative class of methods is policy gradient algorithms, which optimize policies by directly estimating the gradient of expected returns, but they often suffer from high variance and sample inefficiency~\cite{sutton1999policygradient, grondman2012survey, silver2014deterministic}. These issues are particularly acute in environments with large, discrete action spaces. For our task, which involves discrete and computationally expensive transformation evaluations, these drawbacks are especially problematic. Consequently, policy gradient methods are less appropriate for PerfLLM, where stable and sample-efficient learning is essential.

\paragraph{Max Q-Learning}

\begin{figure}
\centering
\includegraphics[width=\linewidth]{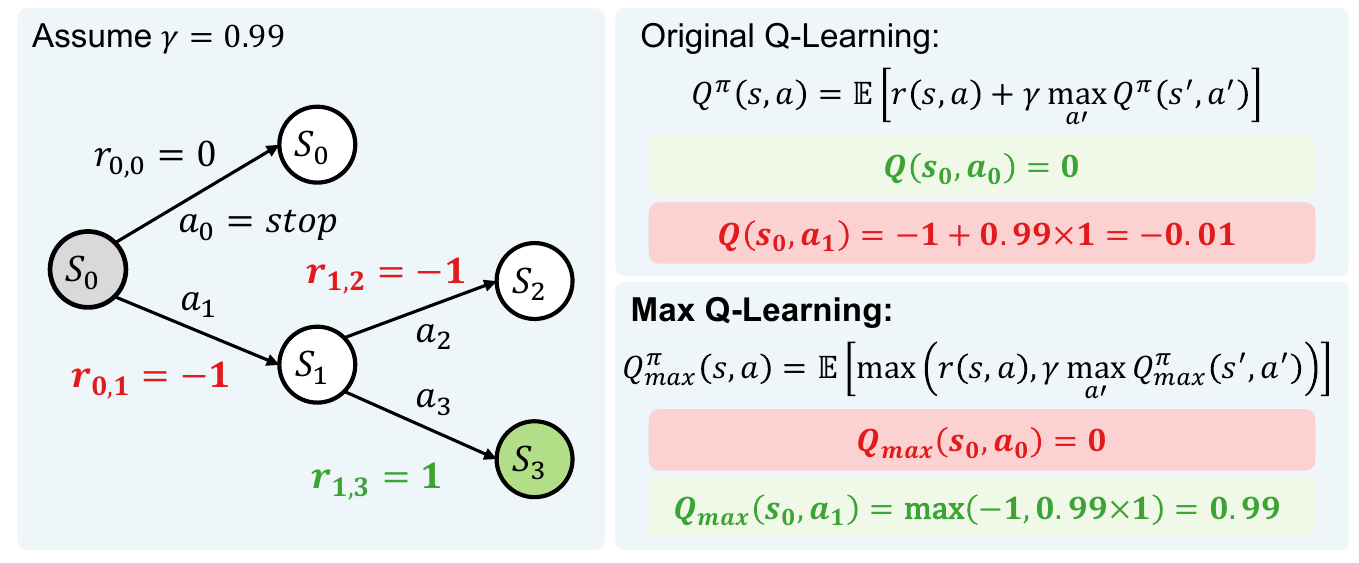}
\caption{An example comparing the Q-value updates in original Q-Learning and Max Q-Learning. The best achievable state $S_3$, highlighted in green, demonstrates how Max Q-Learning explicitly prioritizes trajectories leading to higher peak rewards, thereby selecting action $a_1$, whereas Original Q-Learning selects the immediate "stop" action $a_0$.}
\label{fig:max-q-learning}
\end{figure}

Through experimentation, we found that using the original Q-learning objective was not effective in our scenario because traditional Q-learning is designed to maximize the expected cumulative reward, averaging outcomes across many trajectories. In our context, however, the goal is to identify the single best trajectory that yields the highest possible reward, making standard Q-learning unsuitable. To better address this, we adopted the Max Q-learning approach proposed by Gottipati et al.~\cite{gottipati2023maxq}, which directly focuses on maximizing the best achievable reward within an episode. This modification introduces a revised Bellman equation known as the \emph{max-Bellman} equation:
\begin{equation}
Q_{\max}^\pi(s, a) 
= \mathbb{E}
\Bigl[
  \max(
    r(s,a),\, 
    \gamma \, Q_{\max}^\pi(s', a')
  )
\Bigr].
\end{equation}
\Cref{fig:max-q-learning} illustrates a simple example highlighting this difference. Starting from state \(S_0\), the traditional Q-learning objective would select action \(a_0\) (to stop immediately) due to its higher expected cumulative reward. However, Max Q-learning evaluates action \(a_1\) (leading to state \(S_3\)) more favorably because it explicitly prioritizes trajectories that achieve higher peak rewards. Consequently, in our PerfLLM implementation, we utilize the max-Bellman equation to train the RL agent, guiding it toward discovering and applying transformations that lead to the most optimal code performance.

\subsection{RL Training Techniques}

Through extensive hyperparameter tuning and empirical evaluation, we identified and applied the following set of techniques to enhance RL agent training. Other commonly employed methods in reinforcement learning literature, such as prioritized experience replay~\cite{schaul2016prioritized} and noisy networks~\cite{fortunato2019noisynetwork}, were excluded, as our experiments indicated that these approaches generally did not provide meaningful performance gains in our specific setting.

\paragraph{Experience Replay}

Experience replay is an essential technique for overcoming two challenges in training deep Q-networks~\cite{mnih2013dqn}. First, it enhances sample efficiency by storing each transition in a replay buffer, allowing the network to update its parameters multiple times using the same experience rather than discarding it after a single use. Secondly, it combats the instability caused by the high correlation of sequentially collected data. By randomly sampling mini-batches from the replay buffer, experience replay breaks the temporal correlations between samples, leading to more stable gradient updates and preventing oscillations or divergence in the network parameters.

\paragraph{Double DQN}

Double DQN, introduced by van Hasselt et al. \cite{vanhasselt2015deep}, improves upon standard DQN by addressing training instability and overestimation of Q-values. In traditional DQN, the target is computed as
\begin{equation}
Q(a,s) = r(s,a)  + \gamma \max_{a'} Q(s', a'; \theta),
\end{equation}
where $\theta$ represents the parameters of the policy network, using the same network for both action selection and evaluation, which can lead to unstable updates. Double DQN decouples these by selecting the best action using the current network and evaluating it with a separate target network, whose parameters are denoted as $\theta^-$, yielding the expression
\begin{equation}
Q(s, a)^{\text{DoubleDQN}} = r(s,a) + \gamma Q(s', \argmax_{a} Q(s', a; \theta); \theta^-).
\end{equation}
This approach stabilizes training and reduces bias, leading to faster, more reliable convergence~\cite{mnih2015human}.

\paragraph{Dueling Network}

The dueling network architecture enhances deep Q-learning by decomposing the Q-value function into two distinct streams: one that estimates the state value \(V(s)\) and another that calculates the advantage \(A(s,a)\) for each action~\cite{wang2016duelingdqn}. Instead of directly predicting \(Q(s,a)\), the network computes these components separately and then combines them to obtain the final Q-values. This separation allows the model to learn which states are valuable without needing to learn the effect of each action in every state, which is particularly useful in environments where many actions yield similar outcomes. By isolating the value of a state from the advantage of individual actions, the dueling network architecture improves sample efficiency, stabilizes learning, and often leads to faster convergence in complex scenarios.



\section{Winning the Performance Game}
\label{sec:evaluation}

\begin{figure*}
    \centering
    \includegraphics[width=1.0\linewidth]{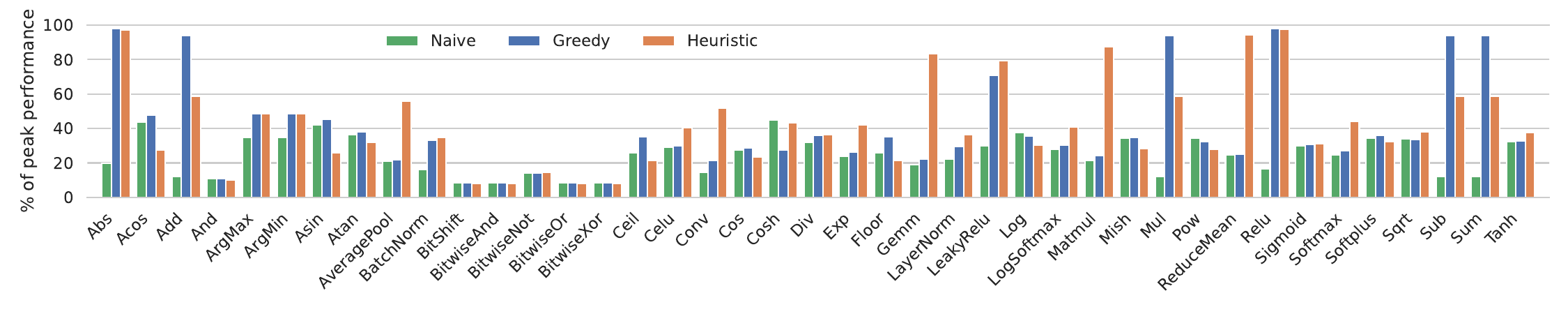}
    \caption{Comparison of micro-kernel performance achieved through passes implementing various transformation strategies for the Snitch RISC-V extensions. \emph{Naive} applies loop fusion and memory reuse until exhaustion. \emph{Greedy} extends the \emph{naive} pass with hardware-specific transformations. \emph{Heuristic} is implemented by a hardware expert as a function that accounts for the structure of the program.}
    \label{fig:microkernels-automatic}
\end{figure*}

\begin{figure}
    \centering
    \includegraphics[width=1.0\linewidth]{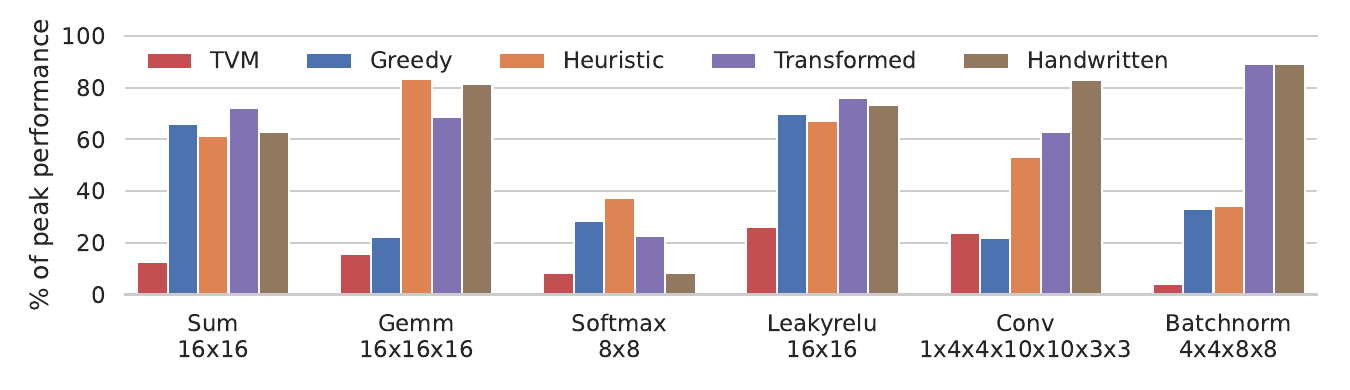}
    \caption{Performance of micro-kernels obtained through automated passes (\emph{greedy}, \emph{heuristic}), manual transformation-centric optimization (\emph{transformed}), TVM 0.11.1, and \emph{handwritten} C and assembly for the Snitch RISC-V extensions.}
    \label{fig:microkernels}
\end{figure}

We begin our evaluation by demonstrating that heuristic-based optimization implemented in PerfDojo is time-competitive with existing methods. In \Cref{sec:snitch-performance}, we use simulation to evaluate its support for novel hardware by targeting an experimental RISC-V design with custom extensions. Subsequently, in \Cref{sec:trad-search}, we show that our heuristic-driven search on conventional hardware outperforms state-of-the-art approaches.

Following this, \Cref{sec:perfllm-eval} evaluates our primary contribution, PerfLLM (\Cref{sec:perfllm}), which discovers high-performance implementations without relying on hardware-specific heuristics to guide the search. In this approach, hardware knowledge is exposed to the search algorithm only as a library of transformations, without any a priori information on their expected performance impact.

\subsection{Heuristic pass for RISC-V}
\label{sec:snitch-performance}

We streamline the process of performance optimization on novel hardware, demonstrating our approach by leveraging the capabilities of the Snitch architecture~\cite{zaruba2020snitch}. Snitch extends~\cite{schuiki2020stream} the open RISC-V ISA and delivers $2\times$ higher energy efficiency in floating-point computations compared to CPUs~\cite{cavalcante2019ara} and GPUs~\cite{ditty2018nvidia} of the same generation. The energy efficiency of the Snitch~\cite{zaruba2020snitch} architecture is achieved by two RISC-V ISA extensions: stream semantic registers~(SSR)~\cite{schuiki2020stream} and floating-point repetition~(FREP).

We conduct our experiments using deterministic cycle-accurate simulation with the Verilator model of Snitch-cluster. We compare the achieved performance against the theoretical compute peak. In most cases, the kernels we analyze involve either floating-point or integer arithmetic exclusively. Therefore, estimating the peak achievable instructions per cycle as 1.0, we calculate the theoretical peak by counting the number of required arithmetic operations.

In \Cref{fig:microkernels-automatic}, we examine three strategies applied during an optimization pass to explore the transformation space. The \emph{naive} strategy imitates the programmer's actions without extensive architectural insight, aiming only to merge scopes and reuse buffers as much as possible. The \emph{greedy} approach incorporates hardware-aware transformations and applies them exhaustively, assuming that their usage is always beneficial. The geometric mean speedup over the \emph{naive} is 46\%.

The \emph{heuristic} strategy leverages hardware expertise by analyzing elements within the IR to select transformations that optimize performance. For instance, kernels optimized using the \emph{greedy} strategy typically achieve only about 25\% of peak performance due to a 4-cycle instruction pipeline latency. To overcome this bottleneck, the \emph{heuristic} strategy seeks to mitigate latency by tiling the outermost loops in each loop nest to a size of 4, whenever possible. For example, if the initial loop nest has dimensions \Verb|[N,D1,D2]| (from outermost to innermost), the transformation reshapes it to \Verb|[N/4,4,D1,D2]|. The dimension of size 4 is then repositioned to the innermost part of the loop hierarchy, resulting in a structure of \Verb|[N/4,D1,D2,4]|. Finally, unrolling is applied to this innermost dimension. This \emph{heuristic} strategy yields a geometric mean speedup of 58\% over the \emph{naive}.

In \Cref{fig:microkernels}, we evaluate the performance of micro-kernel implementations. We compare implementations generated through transformations with those produced using TVM and manually crafted by Snitch cluster developers. Handwritten implementations heavily utilize inline assembly to leverage Snitch features. TVM is provided only as a reference since it does not consider Snitch features, limiting its optimization capabilities. The geometric mean speedup of \emph{transformed} over \emph{handwritten} implementations is 13\%.

Our approach is particularly advantageous for quickly examining potential transformations, as memory accesses are automatically handled by the code generator. This task is typically challenging, especially when configuring SSR and FREP extensions. Our streamlined transformation management enables faster identification of loops that can utilize these extensions. Once such loops are identified, users of this transformation-centric pipeline do not need assembly knowledge to enable these features. Finally, our approach supports applying transformations exhaustively, aiding in uncovering optimization opportunities that may not be immediately apparent.

\begin{figure}
    \centering
    \includegraphics[width=1.0\linewidth]{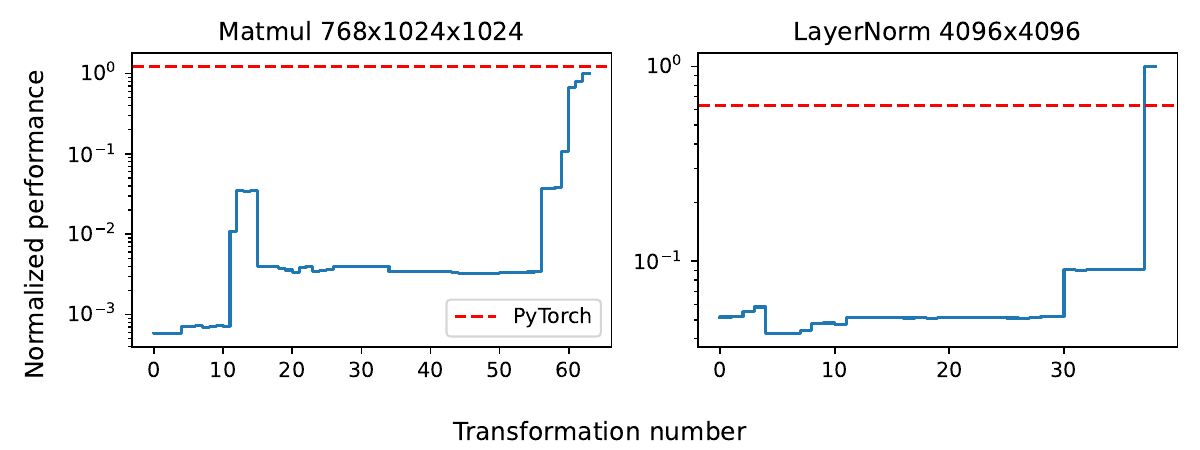}
    \caption{Performance during manual code transformation process.}
    \label{fig:transform-impact}
\end{figure}

\begin{figure*}
    \centering
    \includegraphics[width=1.0\linewidth]{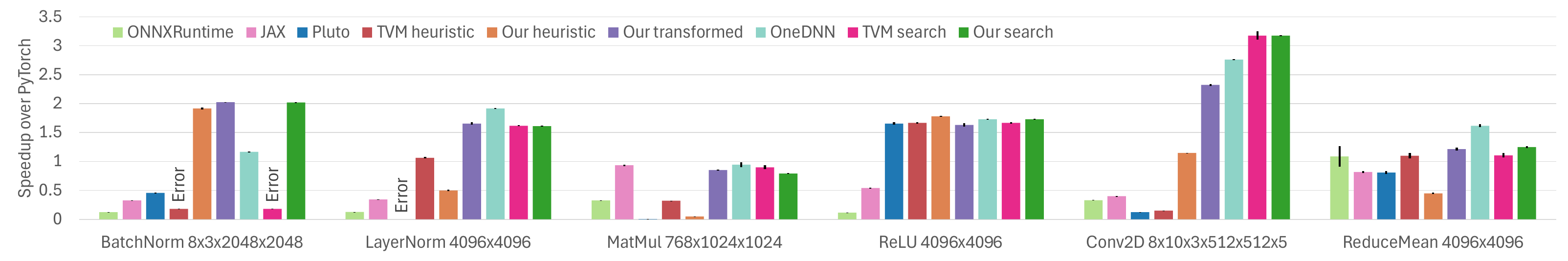}
    \caption{Kernel performance across various frameworks and libraries on x86. The \emph{heuristic} version performs a single program evaluation pass, whereas the \emph{search} version continues until reaching a limit of 1000 evaluations, after which the search terminates. The \emph{transformed} version is derived by applying transformations manually.}
    \label{fig:figure_perf_libraries}
\end{figure*}

\begin{figure*}
    \centering
    \includegraphics[width=1.0\linewidth]{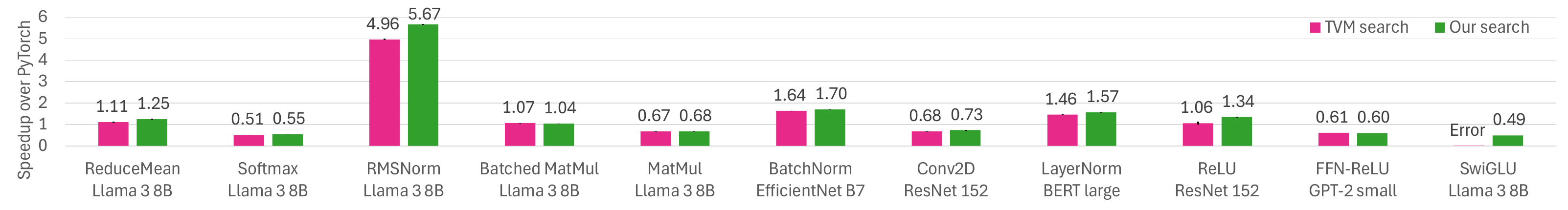}
    \caption{Kernel performance across various shapes in existing models after 1000 auto-tuning evaluations on x86. Excluding SwiGLU, where the TVM auto-scheduler fails to produce a valid schedule, our approach achieves a 7.6\% geometric mean speedup over it.}
    \label{fig:fig_perf_models}
\end{figure*}

\subsection{Heuristic search}
\label{sec:trad-search}

Observing the manual transformation process, illustrated by the performance impact of each transformation in \Cref{fig:transform-impact}, we encountered challenges in using traditional search methods such as greedy search and simulated annealing. These challenges include traversing and escaping local minima, as well as navigating large plateaus of equivalent performance due to transformations that do not immediately affect performance but enable critical optimizations later in the process.

\begin{figure}
    \centering
    \includegraphics[width=1.0\linewidth]{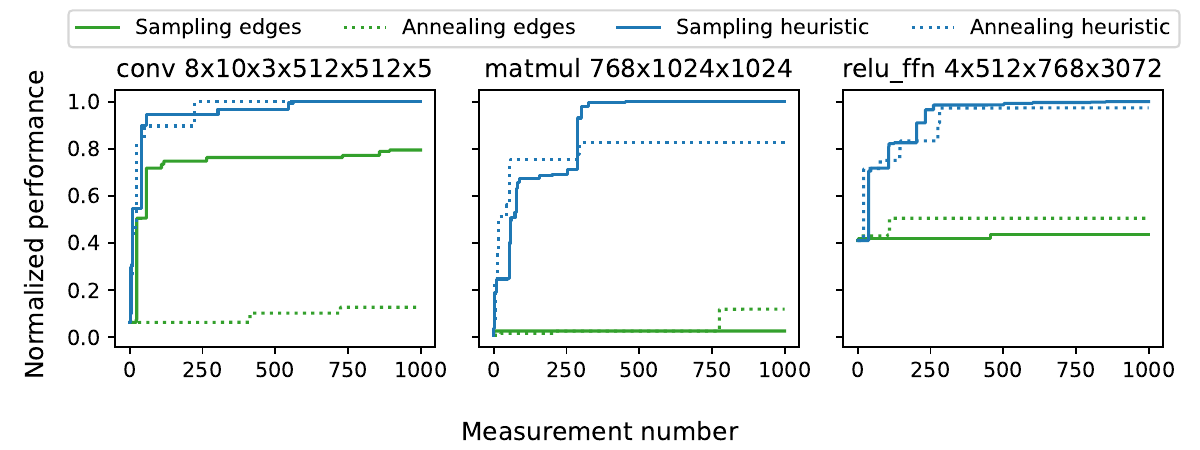}
    \caption{Comparison of convergence speed between simulated annealing and random sampling across differently structured search spaces, based on \emph{edges} within the transformation graph or \emph{heuristic} transitions among transformation candidates.}
    \label{fig:all-search-methods}
\end{figure}

\subsubsection{Structure of the Search Graph}
Initially, we considered a straightforward approach to structuring the search graph by aligning it with the transformation graph. In this setup, the structure of the search graph mirrors the edges of the transformation graph, which we refer to as \emph{edges}-based.

To address the challenges of navigating the search space, we propose an alternative \textit{heuristic}-based method inspired by the expert hand-tuning process. In this approach, the transformation sequence is not constructed sequentially from start to finish. Instead, an initial complete sequence is generated as a candidate and then iteratively refined by modifying selected transformations at arbitrary points, leaving other transformations unaffected. To support this search process, we define a heuristic function specific to the target hardware, which suggests alternative candidate sequences for a given transformation sequence.

\subsubsection{Search Method}
For the search, we implement two strategies that differ in their transformation selection process and cost definition. The first strategy is a global random \emph{sampling} over all previously encountered programs, with selection probabilities based on the costs of past evaluations. Here, we define the cost of a transformation sequence as the runtime of its parent in the search graph. This approach avoids allocating time budget to evaluate children of weakly performing candidates, in contrast to defining a program’s cost as its own runtime. The second strategy relies on simulated \emph{annealing}, which, unlike the first method, allows us to define a program's cost directly as its runtime, thus avoiding the limitations of the sampling-based approach.

In \Cref{fig:all-search-methods}, we compare various search methods and search space structures. The use of heuristics emerges as a decisive factor in the speed of performance convergence, highlighting the importance of supplying heuristics incorporating expert knowledge about hardware specifics.

\subsubsection{Performance on x86}

Evaluation of our heuristic search was performed on all 18 cores of an Intel Xeon CPU E5-2695 v4 with hyper-threading disabled. The following software versions were used: PyTorch 2.3.1, ONNXRuntime 1.18.1, JAX 0.4.31, OneDNN 3.5.3, TVM 0.16.0, Pluto~\cite{bondhugula2008pluto} 0.12.0, and Clang/LLVM 18.1.8. Each evaluation incorporated warmup iterations, with each iteration using new input data. All kernels from PyTorch and JAX utilized their respective just-in-time (JIT) capabilities.
ONNXRuntime evaluation employed its default execution provider. We use the recommended \Verb|--parallel --tile| flags for Pluto with default tile sizes. 

\Cref{fig:figure_perf_libraries,fig:fig_perf_models} demonstrate that the introduced abstractions do not limit attainable performance and remain competitive with both highly optimized handwritten libraries and state-of-the-art auto-tuning on a well-studied CPU architecture.

When using sizes derived from existing models (\Cref{fig:fig_perf_models}), the performance gains from auto-tuning are not consistently superior to those of PyTorch, particularly evident in the convolution kernel. However, with less common sizes (\Cref{fig:figure_perf_libraries}), auto-tuning can surpass handwritten libraries, suggesting limited optimization for these uncommon sizes. This highlights an advantage of the transformation-centric approach over the library-centric approach, as it provides flexibility for discovering more efficient implementations, even on existing architectures.

The results in \Cref{fig:fig_perf_models} demonstrate that our method's performance is frequently comparable to that of TVM. We attribute this to both approaches targeting similar computational and memory bandwidth thresholds while leveraging the same robust LLVM backend, which features well-tuned heuristics for the evaluated architecture. A greater performance disparity is evident on architectures for which the compiler heuristics are less optimized, as exemplified in \Cref{sec:perfllm-eval} (\Cref{fig:speedup-gh200-cpu}).


We observed several cases where existing frameworks struggled with schedule validation. For kernels such as BatchNorm and SwiGLU, after 1000 auto-tuning iterations, the TVM auto-scheduler produced no valid schedules. Attempts were either rejected for exceeding a 15-second compilation timeout or terminated with a runtime error. More critically, Pluto's optimization of the LayerNorm kernel failed numerical validation. These challenges highlight the need for robust correctness guarantees in optimization frameworks.

\subsection{Search with PerfLLM}
\label{sec:perfllm-eval}

\begin{figure}
    \centering
    \includegraphics[width=1\linewidth]{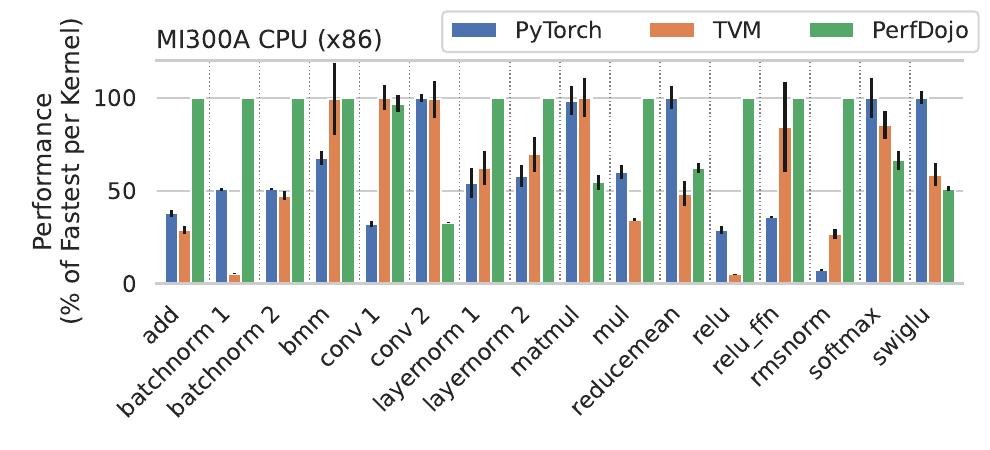}
        \caption{PerfDojo performance compared to PyTorch and TVM. The geometric mean speedup of kernels optimized with PerfDojo on MI300A is $1.56\times$ over PyTorch and $1.80\times$ over TVM.}
        \label{fig:speedup-mi300-cpu}
\end{figure}

We evaluate the performance of PerfLLM on two accelerators: the AMD MI300A and the Nvidia GH200. The following software versions were used: GCC 7.5.0, Clang 19.1.3, CUDA 12.6, ROCm 6.2.4, PyTorch 2.6.0, and TVM 0.19.0.

\begin{table}
\caption{ML operators optimized using PerfLLM.}
\small
\begin{center}
\begin{tabular}
{@{}llr@{}}
\toprule
Label & Input shape & Description \\
\midrule
add & 3072$\times$4096 & Elementwise addition \\
batchnorm 1 & 8$\times$3$\times$2048$\times$2048 & Batch Normalization \\
batchnorm 2 & 8$\times$64$\times$300$\times$300 & Batch Normalization \\
bmm & 192$\times$256$\times$128$\times$256 & Batched Matrix Multiplication \\
conv 1 & 8$\times$10$\times$3$\times$512$\times$512$\times$5 & 2D Convolution \\
conv 2 & 8$\times$64$\times$64$\times$56$\times$56$\times$3 & 2D convolution \\
layernorm 1 & 16384$\times$1024 & Layer Normalization \\
layernorm 2 & 4096$\times$4096 & Layer Normalization \\
matmul & 768$\times$1024$\times$1024 & Matrix Multiplication \\
mul & 6$\times$14336 & Elementwise multiplication \\
reducemean & 4096$\times$4096 & Average along axis \\
relu & 4096$\times$4096 & Rectified Linear Unit (ReLU) \\
relu\_ffn & 8$\times$64$\times$112$\times$112 & ReLU+FeedForward Network \\
rmsnorm & 3072$\times$4096 & Root Mean Square Normalization \\
softmax & 24576$\times$512 & Softmax \\
swiglu & 1$\times$256$\times$4096$\times$448 & SwiGLU activation function \\
\bottomrule
\end{tabular}
\end{center}
\label{tab:kernels}
\end{table}

We present the search results of PerfLLM across a set of deep learning kernels (\Cref{tab:kernels}) in \Cref{fig:speedup-gh200-cpu,fig:speedup-mi300-cpu}. For a significant portion of the considered kernels, the TVM search was unable to identify any valid schedules; consequently, in these instances, we had to use the default schedule. This was primarily due to exceeding the runtime timeout, which TVM defaults to 10 seconds, a limit we also used in our search. This is not unusual behavior and has been frequently reported by TVM users (\cite{tb0,tb1,tb2,tb3,tb4,tb5}). Less frequently, schedules were rejected due to compilation timeouts, often caused by large unrolling factors chosen by TVM. Finally, in a few rare instances, kernels generated by TVM failed numerical verification. Since TVM does not perform this verification itself, assuming all generated schedules are correct, we believe these failures are software bugs rather than a fundamental limitation.

The search process with PerfLLM is more computationally expensive than heuristic-guided methods (\Cref{sec:snitch-performance,sec:trad-search}), resulting in an increase from 5$\times$ to 100$\times$ in runtime. 
For perspective, based on an eight-hour optimization time limit per kernel, we extrapolate that tuning a full library of approximately 160 ONNX operators would require an estimated 1280 node-hours. Nevertheless, this one-time investment represents a substantial saving compared to the engineering effort required to manually achieve a comparable level of performance on new hardware.

Next, we describe the specific kernels discovered by RL that outperform their PyTorch implementations on the GPU.

\begin{figure}
    \centering
    \begin{subfigure}{1\linewidth}
        \centering
        \includegraphics[width=1\linewidth]{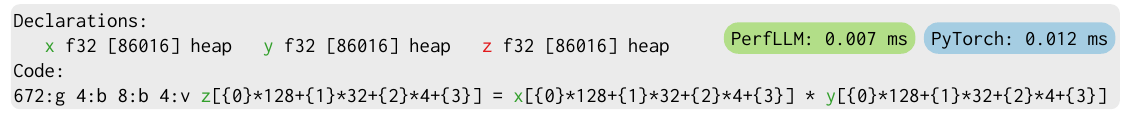}
        \caption{Elementwise multiplication on 86016 elements.}
        \label{fig:gpu-kernels-mul}
    \end{subfigure}
    \begin{subfigure}{1\linewidth}
        \centering
        \includegraphics[width=1\linewidth]{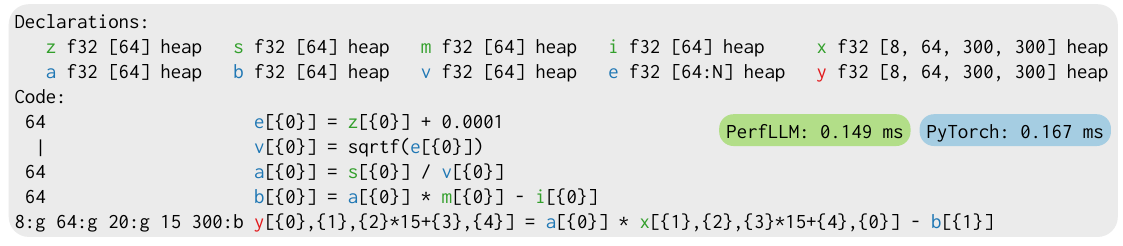}
        \caption{Batch normalization with input size $N=8$ (batch size), $C=64$ (features), and $H=W=300$ (height/width).}
        \label{fig:gpu-kernels-batchnorm}
    \end{subfigure}
    \caption{An overview of the GPU kernel implementations discovered by PerfLLM and their performance.}
    \label{fig:gpu-kernels}
\end{figure}

\paragraph{Elementwise multiplication}

Despite its trivial implementation, the RL-discovered variant outperformed PyTorch by $1.62\times$ and TVM by $1.22\times$ on MI300A, and on GH200 (\Cref{fig:gpu-kernels-mul}), it achieved a $1.71\times$ speedup over PyTorch and $3\times$ over TVM. This demonstrates that RL can identify common optimization techniques used by humans: it vectorized the innermost loop (size 4) to enable 128-bit loads instead of 32-bit, and set the total block size equal to the warp size ($32 = 4 \times 8$), which is a standard practice. We attribute the speedup partly to PyTorch’s use of padding for general input sizes, which reduces efficiency in this specific case.


\paragraph{Batch normalization} On GH200, PerfLLM reached 92\% of PyTorch’s performance, while TVM failed to yield a valid schedule. On MI300A (\Cref{fig:gpu-kernels-batchnorm}), however, PerfLLM’s implementation surpassed PyTorch by 1.12$\times$ and TVM by 1.76$\times$. It achieved this by sequentially computing the temporaries \textcolor{darkblue}{\texttt{e}}, \textcolor{darkblue}{\texttt{v}}, \textcolor{darkblue}{\texttt{a}}, and \textcolor{darkblue}{\texttt{b}} on the CPU before launching the CUDA kernel. Since the input’s height and width are not multiples of the 64-thread wavefront size, both our method and PyTorch incur redundant computation on padded elements. To mitigate this, PerfLLM selected a block size of 300, padding it to 320 by using 5 wavefronts.

\section{Related Work}

Halide~\cite{ragan2013halide} introduced programmable schedules for image pipelines, a foundational idea that influenced the subsequent development of TVM. To overcome the difficulty of integrating highly hardware-specific sketch derivation rules into Ansor~\cite{zheng2020ansor} and to achieve performance comparable to handwritten libraries, a later work~\cite{chen2021bring} proposed embedding problem- and hardware-specific code generation within TVM. Nevertheless, this dependence on a code generator tailored to a narrow problem set limits its applicability to a broader spectrum of user applications.

While TVM has seen significant development, Bolt~\cite{xing2022bolt} demonstrated that Ansor's performance on GPUs reached only approximately 10\% of the efficiency of vendor-optimized cuBLAS. To address this performance gap, Bolt integrated CUTLASS as a configurable micro-kernel. Building upon these efforts, MetaSchedule~\cite{shao2022tensor} was introduced to refine the definition of the schedule space within TVM, aiming to overcome prior limitations. Further extending TVM's framework, TensorIR~\cite{feng2023tensorir} explores and enumerates mappings of low-level iteration blocks to hardware intrinsics, such as Nvidia's Tensor Cores and specialized ARM vector instructions (e.g., sdot). However, this approach demonstrated performance improvements limited to convolution and GEMM kernels.

Triton~\cite{tillet2019triton} specifically targeted programming neural networks on NVIDIA GPUs. It automated parts of scheduling, such as memory coalescing and shared memory management, while leaving users to expose autotunable parameters of kernel implementations. Triton serves as one of the backends for TorchInductor to compile PyTorch~\cite{paszke2019pytorch} models.

Exocompilation~\cite{ikarashi2022exocompilation,ikarashi2025exo} demonstrated the Exo language and transformations, along with the effect analysis framework to detect their applicability, aiming to optimize hardware accelerators in close-to-peak utilization regimes. Due to the high expressiveness of the language, effect analysis is not trivial and may end up being too conservative for optimizing a broader set of kernels than the evaluated GEMM and convolution.

ISA Mapper~\cite{sotoudeh2019isa} explored mapping linear algebra computations to hardware instructions. However, its scheduling space was closer to the macro-kernel level, involving the management of parallelism and communication.

\paragraph{Polyhedral Compilation}

The introduction of the Tensor Comprehensions~\cite{vasilache2019next} IR extended the capabilities of the Halide pipeline to serve as a frontend for deep learning tasks, leveraging the polyhedral model to guide optimizations~\cite{verdoolaege2010isl}. 

Tiramisu~\cite{baghdadi2019tiramisu} is a polyhedral compiler integrated with a scheduling language, leveraging the LLVM infrastructure.

MLIR~\cite{lattner2021mlir} addressed the need to express features of various architectures through domain- and hardware-specific dialects. MLIR was explicitly designed to support polyhedral optimizations. Deep learning frameworks developed by Google, such as TensorFlow~\cite{abadi2016tensorflow} and JAX~\cite{frostig2018compiling}, relied on XLA~\cite{snider2023operator} optimizations implemented using the MLIR framework. MLIR~\cite{lattner2021mlir} provides users with flexibility in customizing optimization passes. Nonetheless, MLIR does not provide an answer to optimization automation.


Overall, a polyhedral IR does not natively support hardware-specific features, making them also unavailable to polyhedral schedulers~\cite{mullapudi2015polymage}. While a polyhedral IR can be extended, for instance, with annotations, such extensions are susceptible to the issues of general-purpose IRs. In particular, the expressiveness of such an IR can make it difficult to either prove semantic preservation or ensure that the range of code implementation variants includes efficient programs that utilize hardware-specific knowledge, especially on novel hardware. In contrast, we start defining an IR with semantic preservation in mind and manage to effectively enable hardware-specific features for a significant portion of the deep learning kernels.
\section{Discussion and Future Work}

Our implementation efforts focused on enabling features that have efficient support from hardware vendors (\Cref{tab:ir-features}). Extending these features is an intriguing direction for future research. Developing a formal theory or classification of these features would help distinguish general control flow from other representational features, providing a clearer foundation for transformation analysis in future frameworks.
\section{Conclusion}

We address the challenge of implementing and optimizing deep learning kernels on modern hardware architectures. Our study focuses on streamlining and automating this process by leveraging program transformations. We propose a transformation-centric workflow that splits the optimization process into two parts: first, defining transformations and detecting their valid application (without altering program semantics); and second, finding effective sequences of transformations for a particular kernel.

This separation enables the application of machine learning methods for identifying transformation sequences, while also allowing engineers to implement custom heuristics as they become familiar with the architecture and develop the ability to make informed estimations for parameters such as tile sizes. However, with the introduction of PerfLLM, we demonstrate that PerfDojo also supports a fully automated, RL–based approach to constructing transformation sequences.



Leveraging the PerfDojo representation, a heuristic pass achieved a 13\% geometric mean gain over hand-optimized RISC-V kernels. On x86, our approach delivered a 7.6\% speedup against a state-of-the-art auto-tuner using only greedy search and simulated annealing. For arm, PerfLLM automated optimization to realize a $6.65\times$ speedup over PyTorch without hardware-specific heuristics. These results demonstrate that semantic-preserving transformations facilitate the automated exploration of transformation sequences, paving the way for developing hardware-specific libraries.


\begin{acks}
This work has received funding from the European High-Performance Computing Joint Undertaking (JU) under grant agreement No. 101034126 (EU-Pilot) and the ERC project PSAP under grant agreement No. 101002047. We thank the CSCS supercomputing center for access to compute resources.

The authors utilized AI assistant tools, including ChatGPT and Gemini, for light editing and proofreading throughout this manuscript. All content and ideas remain the original work of the authors.
\end{acks}

\clearpage

\bibliographystyle{ACM-Reference-Format}
\bibliography{src/bibliography}

\end{document}